\documentclass[11pt,titlepage]{article}
\usepackage{epsfig,graphicx}
\usepackage[margin=1 in]{geometry}
\usepackage{dcolumn}
\usepackage{bm}
\usepackage[normalem]{ulem}
\usepackage{amsmath}
\usepackage{amssymb}
\usepackage{float}
\usepackage{comment}
\usepackage{authblk}
\usepackage[round,numbers,sort&compress]{natbib} 
\usepackage{color}
\title{Free Energy Landscape and Characteristic Forces for the Initiation of DNA Unzipping}

\author[*,$\flat$]{Ahmet Mentes}
\author[$\P$ $\flat$]{Ana Maria Florescu}
\author[$\dag$]{Elizbeth Brunk}
\author[$\ddag$]{Jeff Wereszczynski}
\author[$\sharp$]{Marc Joyeux}
\author[*, $\S$]{ Ioan Andricioaei}
\affil[*]{Department of Chemistry, University of California, Irvine, CA 92697}
\affil[$\P$]{Max-Planck Institute for the Physics of Complex Systems, Dresden, Germany, and Interdisciplinary Research Institute, Universit\'{e} des Sciences et des Technologies de Lille (USTL), CNRS USR 3078, 50, Avenue Halley, 59568 Villeneuve d'Ascq, France}
\affil[$\dag$]{Fuels Synthesis Division, Joint BioEnergy Institute, Emeryville, CA 94608, and Department of Chemical and Biomolecular Engineering, Department of Bioengineering, University of California, Berkeley, CA 94720}
\affil[$\ddag$]{Department of Physics, Illinois Institute of Technology, 3300 South Federal Street, Chicago, IL 60616}
\affil[$\sharp$]{Laboratoire Interdisciplinaire de Physique, Universit\'{e} Joseph Fourier- Grenoble 1, BP 87, 38402 St Martin d'Heres, France}
\affil[$\flat$]{equal contribution}
\affil[$\S$]{\tt{Email: andricio@uci.edu}}

\date{\today}

\begin{document}

\maketitle

\abstract{DNA unzipping, the separation of its double helix into single strands, is crucial in modulating a host of genetic processes. Although the large-scale separation of double-stranded DNA has been studied with a variety of theoretical and experimental techniques, the minute details of the very first steps of unzipping are still unclear. Here, we use atomistic molecular dynamics (MD) simulations, coarse-grained simulations and a statistical-mechanical model to study the initiation of DNA unzipping by an external force.  The calculation of the potential of mean force profiles for the initial separation of the first few terminal base pairs in a DNA oligomer reveal that forces ranging between 130 and 230 pN are needed to disrupt the first base pair, values of an order of magnitude larger than those needed to disrupt base pairs in partially unzipped DNA. The force peak  has an ``echo," of approximately 50 pN, at the distance that unzips the second base pair. We show that the high peak needed to initiate unzipping derives from a free energy basin that is distinct from the basins of subsequent base pairs because of entropic contributions and we highlight the microscopic origin of the peak.} Our results  suggest a new window of exploration for single molecule  experiments.

\vspace{3cm}

\clearpage

\markright{Initiation of DNA mechanical unzipping}

\section{Introduction} 

Many essential genetic processes, such as replication, transcription, recombination and DNA repair, involve unzipping of double-stranded DNA (dsDNA) by proteins that disrupt the hydrogen bonds between complementary bases on opposite strands \cite{Saenger84}. The detailed understanding of the nature of DNA mechanical separation dynamics, and of the energetics and forces for the conformations that occur during unzipping  is also relevant for single-molecule DNA sequencing; high-resolution measurements of the forces may lead to novel ways of sequencing DNA by providing the ability to read the base identities from the distinct signatures resulting when separating the different types of base pairs \cite{McNallyWM08}. Moreover, single molecule studies in which DNA is being pulled on via atomic force microscopy (AFM) \cite{BolandR95}, by optical \cite{bustamante2000} or magnetical \cite{StrickDCDABC03} tweezers, or is unzipped through nanometer-sized pores \cite{meller} are particularly useful to gauge the mechanical response to external stimuli.  Such insights into DNA elasticity \cite{HirshTLLAP13} and the resilience of its double strand to unzipping can provide useful information for the design of nanomechanical devices constructed of DNA \cite{CheMao} and for the build-up of molecularly engineered DNA scaffolds for molecular-size electronics or for crystalline-state biomolecules that otherwise would be impossible to crystalize \cite{Seeman}.

Recent experiments performed by pulling dsDNA apart with a constant force \cite{danilowicz,danilowicz2, weeks, essevazroulet, rief}  show that  dsDNA separates into ssDNA (single-stranded DNA)  when the applied force exceeds a critical value $F_{c} \sim 12$  pN. Moreover, for forces near $F_{c}$, the dynamics of the unzipping process is highly irregular. Rather than a smooth time evolution, the position of the unzipping fork progresses through a series of long pauses separated by rapid bursts of unzipping  \cite{danilowicz2}.  However, because of their low spatial resolution, single molecule techniques cannot yet reveal the first steps of opening a fully base-paired double helix from a blunt end, e.g., the opening of the terminal base pair. For AFM, for example, typical force constants of the cantilever are in the $10-20$ pN/\AA~ range, which, using equipartition arguments, yields fluctuations on the scale of several \AA~ and the best resolution that can be reached via AFM is currently estimated to be on the order of 10 base pairs \cite{albrecht,strunz, krautbauer}. As a consequence, unzipping straightforwardly only the first few base pairs of the sequence can not be achieved in current pulling experiments.

The opening of terminal base-pairs in blunt-end duplexes is important in initiating DNA melting \cite{wong2008, zgarbova2014}. It is also a biologically important step in the action of nucleic acid processing enzymes \cite{KornbergBook} and in nucleic acid end recognition by retroviral integrases \cite{katz2011}. Moreover, the DNA replication process is a good example of instances where double strand DNA must be unzipped mechanically by polymerases \cite{AndricioaeiGHK04}. Concomitantly with the understanding that terminal base pair opening is biologically relevant, it is important to note the fact,  well-established experimentally and computationally, that the first base pair frays \textit{naturally}, and that it exists in a relatively fast equilibrium between paired and unpaired or frayed states. While this equilibrium is fast compared to the time scale of the pulling apparatus that could be used in single molecule experiments to probe the unzipping, it is, at the same time, slow relative to the capabilities of all-atom molecular dynamics simulations. The fraying of first base-pairs has been studied in NMR experiments \cite{nonin1995,kochoyan1998}; this provided estimates for the equilibrium and kinetic constants for the paired-frayed conformational transition. Themodynamical data was consistent with the view that frayed states are unfavorable enthalpically due to loss of stacking stabilization, but that they are stable entropically. The experimental estimates for the populations of the frayed state were in the 10-30$\%$ range for CG pairs, and up to 50$\%$ for AT pairs, with ample variance depending on experimental conditions.  A recent simulation study \cite{zgarbova2014} has additionally provided the atomistic details of terminal base-pair fraying. The kinetics of fraying has been also investigated, with experimental reports concluding that this process is faster than 1 ms \cite{leroy1988}, while in a computational study of multiple free base pair spontaneous stacking/unstacking in aqueous solution at 310 K transitions occured on a time-scale of 10 ns \cite{JafilanKHF12}.

In the theoretical arena, the fundamentals of DNA denaturation has been studied since the 1960s and several generations of  models for DNA unzipping  have been developed \cite{sebastian,bhattacharjee, kafri,lubensky,marenduzzo}. Arguably the most popular ones are the Peyrard-Bishop (PB) model \cite{PeyrardB89} and its extension, the Peyrard-Bishop-Dauxois (PBD) model \cite{DauxoisPB93} with an extra term in stacking to better reproduces experimental data.  They have  been extensively used to describe  DNA thermal denaturation \cite{peyrard2004} or the dynamics of pulling DNA  by an external force  \cite{ZdravkovicS06}. Other  models have been developed  to investigate quantitatively the difference between DNA unzipping by force and thermal or fluctuation-induced melting of dsDNA \cite{bhattacharjee}, or for studying  interactions between two single DNA strands \cite{lubensky}. 

These modeling approaches, while revealing the fundamental statistical mechanical picture, are not detailed enough  to capture all intricacies of unzipping. An important advancement has been recently made via a semi-microscopic theory of DNA mechanical unzipping advanced by Cocco \emph {et. al.} \cite{cocco1,cocco2}. This theory accounts for hydrogen bonds, stacking  interactions and elastic forces to investigate experimentally observable aspects of DNA unzipping by externally applied forces or torques. Quite interestingly, the calculation of the forces needed to keep  the two extremities of the dsDNA molecule separated by a given distance lead in this model to the prediction for the existence of an extremely large force barrier that opposes initial double helix unzipping, namely a $\sim 250$ pN  force peak occuring at $\sim 2$ \AA~separation from the equilibrium base-pair distance \cite{cocco2}. This is remarkable because it is more than an order of magnitude larger than the unzipping forces for DNA in the bulk (i.e., forces averaged over scores of unzipped basepairs) previously measured in the various experimental settings. 

Because both analytically solvable models and experiments can only reveal a limited number of observables (e.g. force and extension), it is crucial that they are complemented by all-atom simulations. This allows one to better understand the dynamics and observe the microscopic effects that pulling forces have on all degrees of freedom and physical properties of the system \cite{MaffeoYCWLA14}. A previous atomistic molecular dynamics study of DNA mechanical denaturation \cite{santosh2009} focused on the sequence effects that occur during non-equilibrium DNA unzipping (with pulling speeds orders of magnitude larger than those in single molecule tweezer experiments) and not on the equilibrium forces needed for the initiation step. The authors observed jumps and pauses in denaturation  which they attribute to the inhomogeneity of the DNA sequence they have used:  AT (adenine-thymine) rich regions melt earlier (that is at smaller forces) than  GC (guanine-cytosine) rich regions because AT base pairs contain two hydrogen bonds whereas GC base pairs contain three hydrogen bonds.
\begin{figure}[H]
   \vspace{0 cm}
\centering{\includegraphics[width=0.8 \textwidth]{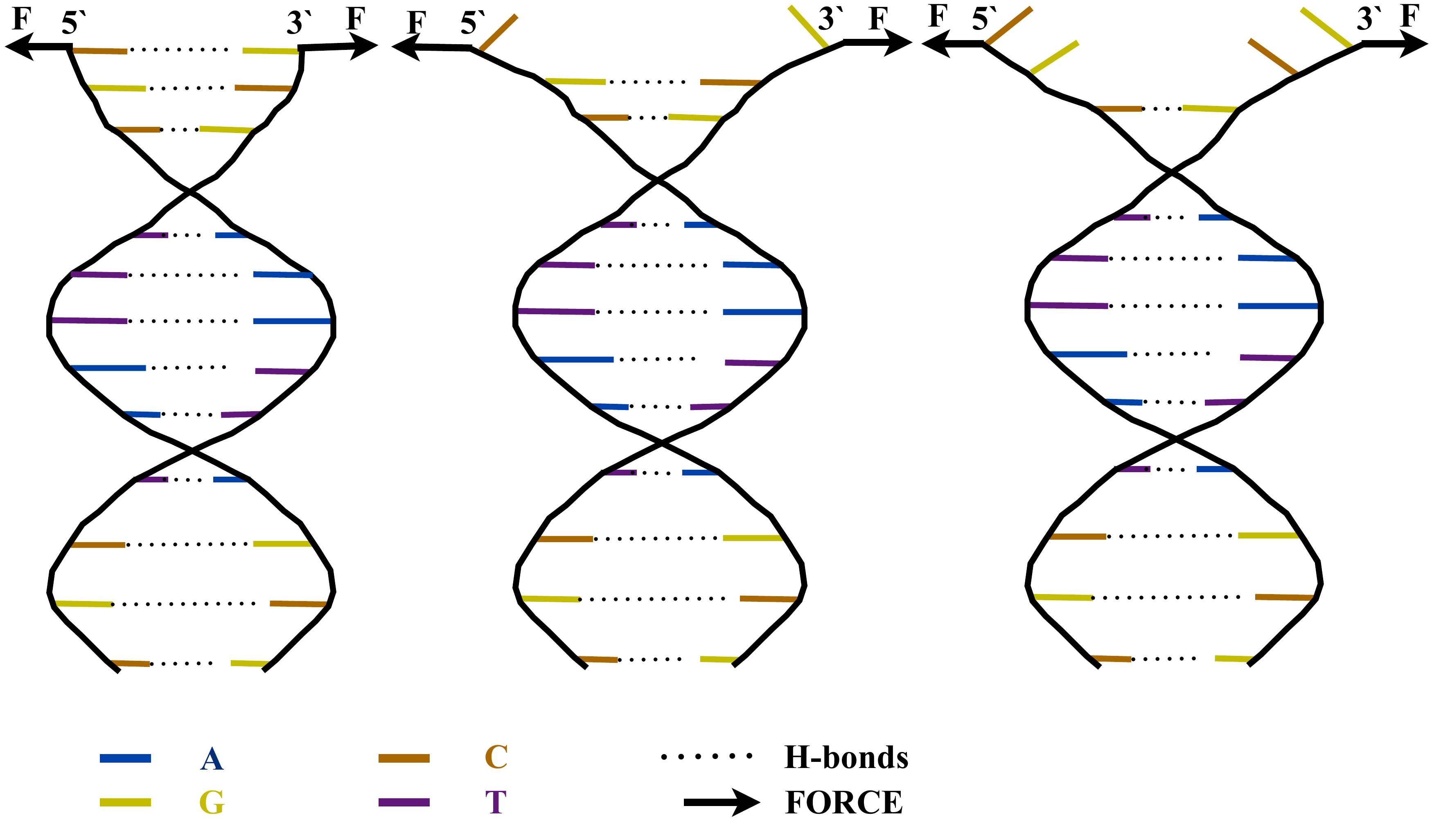}}
\caption{ Schematic representation of the DNA sequence and the forces applied for unzipping: bases $G$, $C$, $A$ and $T$ are shown in yellow, orange, blue and magenta, respectively and the bonds forming the base-pairs are shown with a dotted line. Forces were applied on two backbone atoms of the first base pair: the O$3'$ atom of the $C1$ residue  at the 5' end of one strand and on the  O$3'$ atom of the $G12$ residue at the 3' end of the other strand in the all atom simulations, and on the pseudo-atoms describing the sugars of the same residues in the coarse-grained simulations. The terminal C1-G12 base pair is an example of what we refer to in the text as the ``first" base pair that needs a higher force to be unzipped in comparison to the subsequent base pairs.} 
\vspace{0.2cm}
\label{dnastructure}
\end{figure}

The purpose of the present study is to provide a better understanding of the onset of DNA denaturation and an exploration of the origin of any unusually high forces that occur at the very early stage of unzipping (that is the opening of the first 1-2 base pairs) via detailed molecular simulations and subsequent theoretical analysis. To this end, the rest of the paper is organized as follows. First, we compute, along the base separation coordinate, mean force and  free energy profiles of a dodecamer of helical B-DNA with the base sequence  d(CGCAAATTTCGC)$_2$  using molecular dynamics simulations, umbrella sampling and the weighted histogram analysis method (WHAM) to obtain and atomically detailed potential of mean force profile. Then, in addition to atomistic calculations, to explore the presence of force peaks in simulations with lower (mesoscopic) levels of details and to extend the range of DNA sequences studied, we also perform simulations using a coarse-grained DNA model with three sites per nucleotide \cite{Florescu2011}. Lastly, we also show that we can derive this force peak analytically in the formalism of the Peyrard-Bishop model \cite{PeyrardB89}. Taken collectively, our simulation and analysis results reveal that the opening of the first DNA base pair needs significantly larger forces than the opening of the subsequent ones not only to break the hydrogen bonds that form that base pair, but also to overcome an entropic barrier due to stacking interactions. Additionally we reveal that a`second order' contribution to the force peak stemming from non-native H-bonds (i.e., between base pairs that were not originally H-bonded in the intact dsDNA) exists, and that, concomitantly  to the development of the force peak, a peak in the torque about the DNA axis develops upon initial unzipping.

\section{Simulation Methods}
\subsection{All-Atom Molecular Dynamics Simulations}
We simulated at atomic detail the first steps in the mechanical denaturation of a dodecamer of helical B-DNA   with the sequence  d(CGCAAATTTCGC)$_2$. Fig. \ref{dnastructure} depicts schematically the DNA sequence and the  forces that are exerted to induce mechanical unzipping.; it displays the four nucleotides G, C, A and T and the hydrogen bonds between the base pairs. The same labeling and coloring strategy as in Fig. \ref{dnastructure} is followed for the other figures of this article. The external forces applied to initiate unzipping are also shown. In the atomistic simulations they were applied on the two O$3'$ atoms of the first base pair, i.e., the O$3'$ atom of the C1 residue at the 5' end (i.e., the first Cyt residue on one strand) and the O$3'$ atom of the G12 residue at the 3' end of the other strand. (Numbering is such that  bases are labeled 1 through 12 from the 5' to the 3' direction on both strands, see also Fig. \ref{dna_unzipped}). To account for the natural fraying of terminal base pairs, two starting geometries were used in the simulation: a paired first base pair and a frayed one. For the frayed first base pair case, the initial equilibrium distance between O$3'$ atoms pulled apart is the same as that for the paired case, and the C1 residue at 5' end (on DNAs1-DNA strand1) and the G12 residue at 3' end (on DNAs2- DNA strand2) were flipped out of the backbone by alterations of the corresponding dihedral angles. The structures such prepared then underwent molecular dynamics simulations using version 34 of the CHARMM software package \cite{CHARMM09}, with the CHARMM27 nucleic acid parameters \cite{foloppe, mackerell2}. The reaction coordinate was defined as the separation between the C and G O$3'$ atoms of residues $1$ and  $12$, respectively, and we applied harmonic constraints to the separation distance $\rho$ of these atoms. The functional form of the potential used was $k_u(\rho-\rho_0)^2$. Using umbrella sampling trajectories, statistical data for free energy calculations was collected during the structural changes along the coordinate. The selected atoms were harmonically restrained such as to maintain a separation within approximately $1$ \AA~ of the specified equilibrium distance $\rho_0$, ranging from $14.50$ \AA, which is the base-pairing equilibrium value, to $30.00$ \AA~(which corresponds to 15.5 \AA~separation from base-pairing equilibrium) with increments of $0.05$ \AA~for the first 4.5 \AA~separation and $0.25$ \AA~for the rest of the windows. The last configuration of the trajectory in each window was used as the initial condition for the next window. A total of $130$ windows were calculated, each running for $800$ ps. Each $800$ ps trajectory was then post-processed using WHAM \cite{kumar, grossfield}. The more numerous umbrella sampling windows for the first 4.5 \AA~ we generated for better resolution in the initial free energy profile.  We used the explicit solvent TIP3P potential for water \cite{jorgensen}. The DNA structure was overlaid with a water box previously equilibrated at 300 K, with dimensions 56 \AA~  $\times$ 56 \AA~  $\times$ 56 \AA, and was initially aligned so that the DNA molecule's primary axis is parallel  to the $x$-axis. The water box contained $3362$ TIP$3$P water molecules, and $22$ sodium ions, needed to make the solution electrically neutral. Periodic boundary conditions were used and electrostatic interactions were accounted for using the Particle-Mesh Ewald Method \cite{essmann}, with a real-space cutoff at $12.0$ \AA~ for non-bonded interactions. The leapfrog Verlet algorithm was used with Nos\'{e}-Hoover dynamics \cite{nose, hoover} with a coupling constant (thermal inertia parameter) of 50 internal (AKMA) units \cite{brooks}  to keep the temperature constant at $300$ K throughout the simulations. The system underwent $100$ steps of steepest-descent minimization followed by $1000$ steps of the adaptive-basis Newton-Raphson minimization. It was then heated to $300$ K over an equilibration period of $800$ ps with harmonic restraints applied to the O$3'$ atoms in order to prevent the helical axis from becoming unaligned with the $z$-axis, restraints which are then gradually removed during the production runs. The SHAKE algorithm \cite{ryckaert} was used to constrain all covalently-bound hydrogen atoms.

The biased umbrella sampling trajectories were post-processed with the weighted histogram analysis method (WHAM) \cite{kumar} (as implemented in Ref. \cite{grossfield}), in order to obtain the unbiased free energy values as well as thermodynamic  quantities  from an unbiased system.  Error bars were calculated with Monte Carlo bootstrap error analysis, with repeated computations of the average of resampled data and calculation of the standard deviation of the average of the resampled data. The later is an estimate for the statistical uncertainty of the average computed using the real data.  Because for separation distances below 4.5 \AA~the energy is averaged over more windows than for higher separations, this leads to the smaller error bars in that region of interest.  The force along the reaction coordinate was computed as the derivative of $W(\rho)$ with respect to $\rho$, $F(\rho)=  - dW/d\rho$ and is rigorously \cite{Kirkwood} the canonical-ensemble thermodynamical average of the  force needed to keep the two strands separated by a distance $\rho$.  

To determine the vectorial force components, the mean force along the Cartesian $x$, $y$ and $z$ axes was computed by taking the derivative of $W(\rho)$ with respect to $x_i$, $y_i$ and $z_i$ of the $i$-th atoms involved, 
\begin{equation}
\langle f_{x_i} \rangle \equiv - \langle \frac{dW}{dx_i} \rangle=-\frac{dW}{d\rho} \langle\frac{d\rho}{dx_i}\rangle=-\frac{dW}{d\rho}\frac{ (\langle x_i \rangle - x_{ref})}{\rho\cdot N},
\end{equation}
with the corresponding permuted expressions for the $y$ and $z$ directions.  We were also interested to compute any torque $\tau$ to which DNA is subjected upon increasing the separation distance by finding the backbone vector forces (on atoms C1:O$3'$  and G12:O$3'$) in the $x$, $y$ and $z$ directions, and performing the cross product with the radii vectors of the DNA helix \cite{WereszczynskiA06}. For example, the torque in the helical $z$-direction is
\begin{equation}
\langle \tau_{z_i} \rangle =-\frac{dW}{d\rho} \cdot \frac{ (x \hat{\imath}
+y \hat{\jmath}) \times ((\langle x_i \rangle - x_{ref}) \hat{\imath}
+(\langle y_i \rangle - y_{ref}) \hat{\jmath})}{\rho\cdot N},
\label{torqueeqn}
\end{equation}
with similar expressions for the $x$ and $y$ directions.

\subsection{Coarse-grained simulations}
For coarse-grained simulations we used the three-site-per-nucleotide DNA model developed by Knotts \textit {et al.} \cite{dePablo2007} with the parametrization described in Ref. \cite{Florescu2011}. In this model, each nucleotide is mapped onto three interaction sites (beads): one for the phosphate, one for the sugar ring and one for the base. The equilibrium positions of the three beads are derived from the  coordinates of the atoms  they replace as follows:  for phosphates and sugars, the bead is placed in the center of mass of the atomic structure of the respective group, for adenine and guanine it is placed on the position of the N1 atom of the geiven base, while for cytosine and thymine it has the coordinates of the N3 atom of the given base. 
The interaction potential between these beads comprises six terms:
\begin{equation}
E_{pot}=V_{bond}+V_{angle}+V_{dihedral}+V_{stack}+V_{bp}+V_{qq},
\end{equation}
with $V_{bond}$, $V_{angle}$ and $V_{dihedral}$ the bonded contributions (stretch, angle bending and torsion respectively), and base stacking ($V_{stack}$), base pairing ($V_{bp}$) and electrostatic interactions ($V_{qq}$) --the non-bonded terms-- described by:
\begin{gather}
V_{bond}=k_{1}\sum_{i} (d_{i}-d^{0}_{i})^{2}+k_{2}\sum_{i}(d_{i}-d^{0}_{i})^{4} \nonumber \\
V_{angle}=\frac{k_{\theta}}{2}\sum_{i}(\theta-\theta^{0}_{i})^{2} \nonumber  \\
V_{dihedral} = k_{\phi}\sum_{i}[1-\text{cos}(\phi_{i}-\phi^{0}_{i})] \nonumber \\
V_{stack}=\epsilon\sum_{i<j}\left[\left(\frac{r^{0}_{ij}}{r_{ij}}\right)^{12}-2\left(\frac{r^{0}_{ij}}{r_{ij}}\right)^{6}+1\right] \nonumber \\
V_{bp}=\epsilon_{AT}\sum_{AT~base~pairs}\left[5\left(\frac{r^{0}_{ij}}{r_{ij}}\right)^{12}-6\left(\frac{r^{0}_{ij}}{r_{ij}}\right)^{10}+1\right] + \nonumber \\
+ \epsilon_{GC}\sum_{GC~base~pairs}\left[5\left(\frac{r^{0}_{ij}}{r_{ij}}\right)^{12}-6\left(\frac{r^{0}_{ij}}{r_{ij}}\right)^{10}+1\right] \nonumber \\
V_{qq}=\frac{e^{2}}{4\pi\epsilon_{H_{2}O}}\sum_{i<j}\frac{exp-\left(\frac{r_{ij}}{\kappa_{D}}\right)}{r_{ij}},
\end{gather}
where $d_{i}$ denotes the distance between two beads connected by the bond $i$, $ \theta_{i}$ is the  angle between three consecutive sites on the same strand and $ \phi_{i} $ is the dihedral angle defined by four consecutive beads (also along the same strand). In the non-bonded terms, $r_{i,j}$ is the distance between sites $i$ and $j$. In all equations the values with the superscript index $0$ are equilibrium values for the respective quantities. For their numerical values we refer the reader to reference \cite{dePablo2007}. The force constants for the bonded terms are the following: $k_{1}=0.26$ kcal\textperiodcentered mol$^{-1}$\AA$^{-2}$,  $k_{2}=26$ kcal\textperiodcentered mol$^{-1}$\AA$^{-4}$, $k_{\theta}=104$ kcal\textperiodcentered mol$^{-1}$,  $k_{\phi}=1.04$ kcal\textperiodcentered mol$^{-1}$. The stacking interactions act between the first and second nearest neighbors and $\epsilon = 0.26$ kcal\textperiodcentered mol$^{-1}$. The base pairing term acts  only between native pairs, with  $\epsilon_{AT} = 3.90~ $kcal\textperiodcentered mol $^{-1} $  and $\epsilon_{GC}= 4.37$  kcal\textperiodcentered mol$^{-1}$. Finally, electrostatic interactions are considered to occur only between phosphates, which carry one elementary charge each. In the expression of the Debye-H\"{u}ckel potential  $e$  is the electron charge, $\epsilon_{H_{2}O}= 78\epsilon_{0}$  is the dielectric constant for water at  room temperature expressed as a function of the dielectric permittivity of vacuum, and $r_{D}=13.603$  \AA~  is the Debye length for 50 mM Na$^{+}$ ion concentration.  Compared to Ref.  \cite{dePablo2007}, we use different values $\epsilon_{AT}$ and $\epsilon_{GC}$, and exclude non-native base pairing; this was done because the original set of parameters can sometimes cause the formation of two pairing bonds per base, a fact which induced melting temperatures that were too high. Consequently we modified the pairing energy values to correctly describe thermal and mechanical denaturation. We simulated the mechanical unzipping of the dodecamer simulated in the atomistic model, d(CGCAAATTTCGC)$_2$, plus two other dodecamers, d(TGCAAATTTCGC)$_2$ and d(CTCAAATTTCGC)$_2$ in which we changed the first, and respectively, the second base pair from CG to AT. Dynamics was propagated by integrating Langevin's equations:
\begin{equation}
m_{j}\frac{d^{2}\textbf{r}_{j}}{dt^{2}}=-\nabla E_{pot}-m_{j}\gamma\frac{d\textbf{r}_{j}}{dt}+\sqrt{2m_{j}\gamma k_{B}T}\xi_{j}(t)
\label{Langevin}
\end{equation}
 where $m_{j}$ is the mass of site $j$, $\textbf{r}_{j}$ is its position vector, with the friction coefficient $\gamma$ and the Gaussian white noise $\xi_{j}(t)$ obeying fluctuation-dissipation: 
\begin{equation}
\langle \xi_{i}(t)\xi_{j}(t') \rangle =\delta_{i,j}\delta(t-t')
\end{equation}   
The first term on the right hand side of  eq. \ref{Langevin} denotes the forces resulting from the potential, the second one describes the friction due to the solvent, while the third one is a thermal random noise. We integrated the equations of motion using a second order algorithm with a time step of 10 fs and a friction coefficient $\gamma $ of 5 ns$^{-1}$. A detailed discussion of the choice of $\gamma$ and of its influence on the simulation results can be found in reference \cite{Florescu2011}. The temperature was set to 293 K. We have modeled mechanical unzipping of the DNA sequence by  pulling apart  with a constant rate the  sugar groups that are part of the first base pair. We computed then the average of  the projection along  the separation axis of the internal forces acting on the two beads.  For each point  the force was averaged over $10^{7}$ time steps  (0.1 $\mu$s), corresponding to an increase in separation distance of 0.1 \AA. We have previously used this approach and validated the model and its parameters with respect to DNA unzipping; we refer the reader to Ref. \cite{Florescu2011} for the details of this validation.

\section{Results}
\begin{figure}[H]
\vspace*{0cm}
\centering{\includegraphics[width=0.8\textwidth]{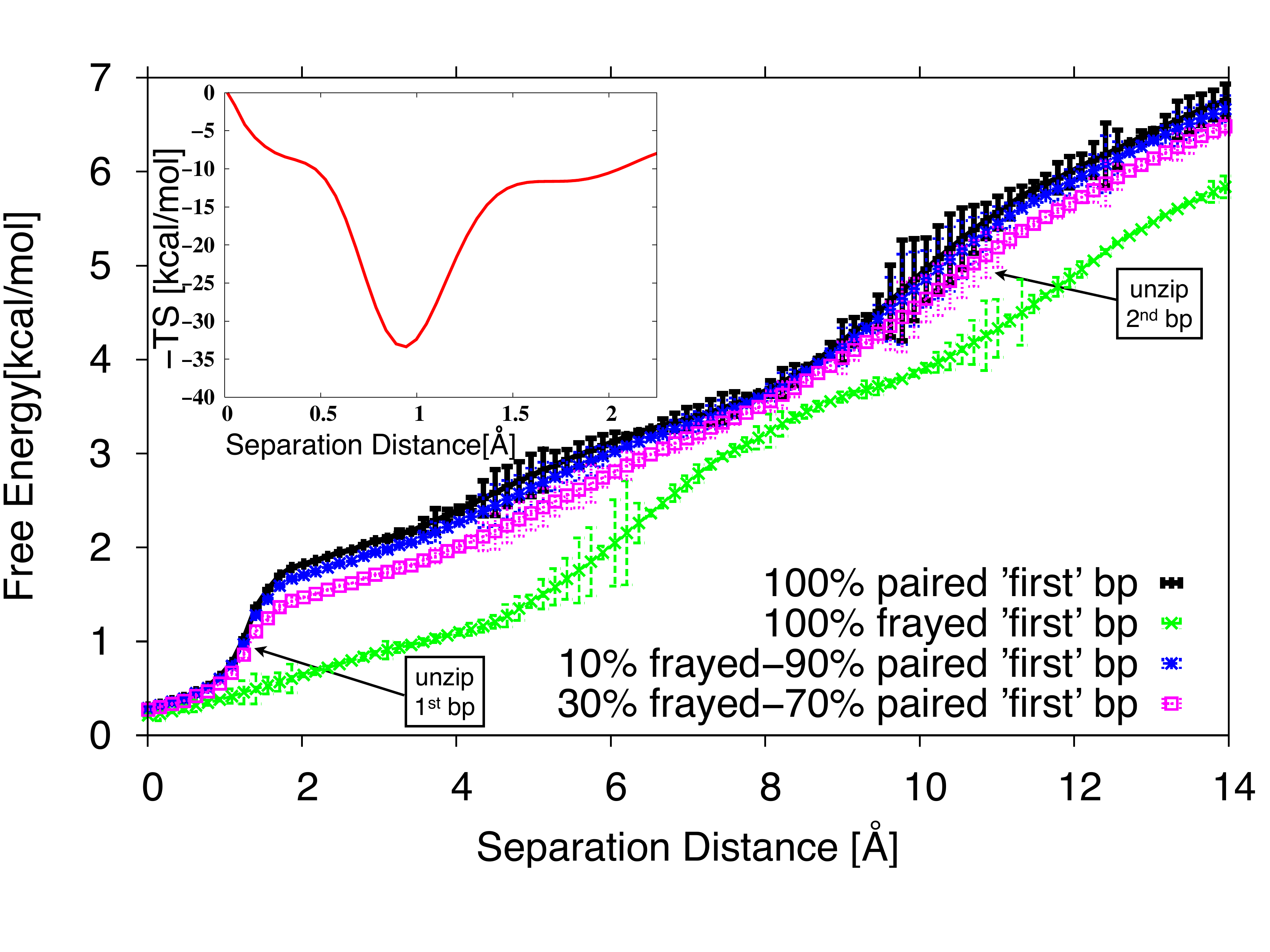}}
\caption{ Free energy of DNA unzipping from all-atom simulations with various frayed$-$paired populations for the first base pair. First base-pair fully paired (black), fully frayed (green), $10\%-90 \%$ frayed$-$paired (blue), $30\%-70\%$ frayed$-$paired (pink). Arrows point to onset of first and the second base pair unzipping. The significantly deeper, ``antlion-pit" free energy well for first base pair separation (between 0-2 \AA) is a key feature explaining a steep increase in the force required for initial unzipping (see main text). Inset: The conformational entropy contribution to the the free energy profile along the separation distance for unzipping the first base-pair.}
\label{FreeEnergy}
\end{figure}

\begin{figure}[H]
\vspace{0.0cm}
\centering{\includegraphics[width=0.8\textwidth]{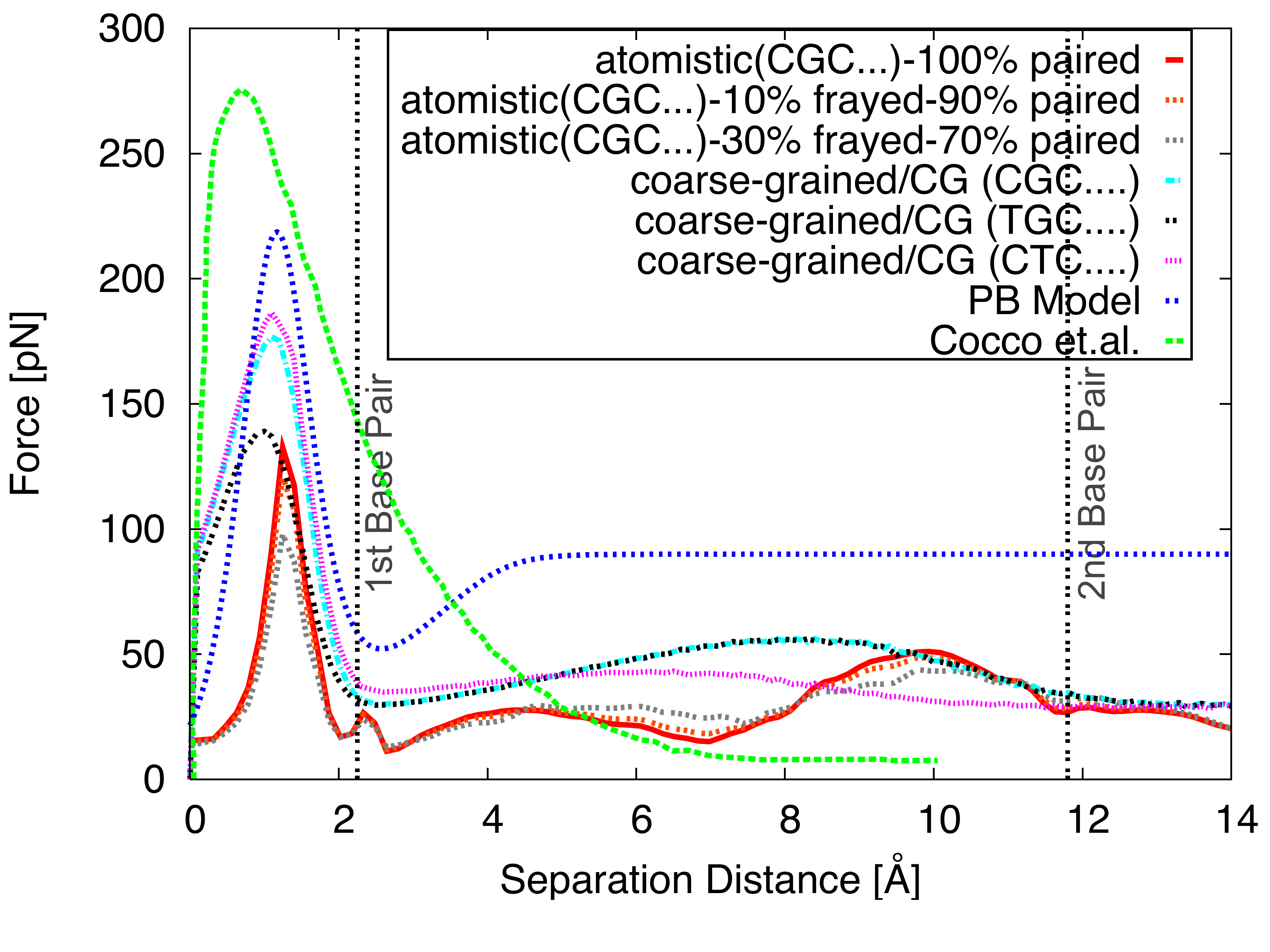}}
\caption{ Mean force on the terminal base pairs need to maintain them at a given separation, as a function of that separation distance. Atomistic simulations in red for fully paired first base-pair in red, in orange dashed line for the case of first base pair 10$\%$ frayed and 90$\%$, light brown dashed for 30$\%$ frayed and 70$\%$ paired; together with the results of Ref. \cite{cocco2} in green dashed line, the Peyrard-Bishop model (dotted dark blue) and the coarse-grained simulations (pink dotted line). We also show (in light blue and black) results for coarse-grained simulations of the same sequence with the first base pair and second base pairs changed from C-G to A-T. The vertical lines mark the distances beyond which the first and second base pairs are fully unzipped. }
\label{Force}
\end{figure}

Using the computational techniques presented in the Simulation Method section, we have performed equilibrium studies of the forces required for the mechanical unzipping of short DNA sequences. We have simulated pulling by the first base pair up to a separation distance of 14 \AA~from equilibrium using all-atom molecular dynamics (CHARMM), and up to 50 \AA~using the coarse-grained model described above. The analysis of the simulation results presented here is mainly focused on what happens at small separation distances, that is, corresponding to the opening of the first two base pairs.  To our knowledge, this is the first time that the onset of DNA mechanical denaturation is studied in such detail. From the all-atom simulations we computed a free energy profile (the potential of mean force (PMF)), whose derivative with respect to the separation coordinate was used, according to the definition of the PMF \cite{Kirkwood}) to obtain the average force needed to keep the first base pair open at any given separation. Because terminal DNA base pairs are known to fray, we had to use two sets of initial conditions, one base-paired, one frayed (as described in the Introduction section). Additionally, since the fraying/unfraying equilibrium is established on time scale much shorter than that of the pulling aparatus used in single molecule pulling (typically A per ms), during a typical pulling one expects to experience time averaging between the two states, so pulling would give the weighted mean (e.g., 30-70\% or 10-90\%) of the fully paired and fully frayed profiles. The free energy profile along the separation coordinate is shown in Fig. \ref{FreeEnergy}. Over a baseline of increasing free energy as a function of separation (whose constant slope averages to the value of the minimum bulk force needed to unzip DNA), we observed significant ``pits'' or free energy. They introduce higher slopes in the profile, and, since the slopes are proportional to the magnitude of the mean force, they are responsible for the  larger forces for the separation of the first and second base pairs (see Figure \ref{Force}). We also computed conformational entropies \cite{kushick} at each separation distance using the quasi-harmonic analysis method \cite{AndricioaeiK01}, in which quasi-harmonic frequencies were calculated from diagonalizing the mass-weigthed covariance matrix of nucleic acid atomic conformational fluctuations in each umbrella sampling window. The calculated conformational entropy profile around the first base pair unzipping is shown in the inset to Fig. \ref{FreeEnergy}. We observed a substantial conformational entropy contribution to the ``ant-lion� pit in the free energy profile along the first base-pair unzipping distance, pointing at a large entropic contribution to the initial slope of the free energy and hence to the force peak.

From the coarse-grained simulations we computed directly the corresponding force at each distance by equilibrium averaging of the force needed to keep the distance between the phosphates of the first base pairs at a given separation. We focus on the analysis of the force peak obtained via these simulations in this section. Before that, we start with an analytical computation of the forces using the Peyrard-Bishop model \cite{singh2005, PeyrardB89}. According to this model the potential energy of a sequence of $N+1$ base pairs whose first base pair ($n=0$) is pulled by a force F perpendicular to the sequence is \cite{singh2005}:
\begin{equation}
V= \sum_{n=0}^{N}[D(1-e^{-ay_{n}})+\frac{K}{2}(y_{n}-y_{n+1})^{2}]-F y_{0}
\label{pb1}
\end{equation}
where $y_{n}$ is the deviation from equilibrium of the distance between the bases of the $n-$th base
pair. The first term describes the pairing interaction and the second one the stacking interaction. By differentiating Eq. \ref{pb1} with respect to $y_{n}$, one obtains:
\begin{gather}
\frac{\partial V}{\partial y_{0}}= 2aD e^{-ay_{0}}(1-e^{-ay_{0}})+K (y_{0}-y_{1})-F \nonumber \\
\frac{\partial V}{\partial y_{n}}= 2aD e^{-ay_{n}}(1-e^{-ay_{n}})+K (2y_{n}-y_{n+1}-y_{n-1})  \nonumber \\
\text{for}~n>0
\label{pb2}
\end{gather}
Imposing that the conformation with minimum energy satisfies $\partial V/\partial y_{n}= 0$ for all values of $n$ in the second equality of Eq. (\ref{pb2}) and taking its continuum $n$ limit, one gets
\begin{equation}
\frac{d^{2} u}{d n^{2}}- A e^{-u}(1-e^{-u}) = 0
\label{pb4}
\end{equation}
where $u = ay$ and $A = 2a^{2} \frac{D}{K}$. The two general solutions of this equation read
\begin{equation}
u = \ln[\frac{A}{C_{1}^{2}}+\frac{1}{4C_{1}} e^{\pm C_{1}(n+C_{2})}+\frac{A}{C_{1}} (\frac{A}{C_{1}^{2}}-1) e^{\mp C_{1}(n+C_{2})}]
\label{pb5}
\end{equation}
where $C_{1}$ and $C_{2}$ are two integration constants. The physical solution requires that $u$ tends towards $0$ when $n$ tends towards $+\infty$, meaning that the far end of the sequence is still zipped, which is possible if and only if $C_{1}^{2} = A$. The physical solution can therefore be recast in the form:
\begin{equation}
u (n_{thresh} \mid n) = \ln [1+\frac{1}{4\sqrt{A}} e^{-\sqrt{A} (n-n_{thresh})}]
\label{pb6}
\end{equation}
where, due to the large value of $A$ (see below), $n_{thresh}$ represents approximately the rank of the base pair up to which the sequence is unzipped. The first line  of Eq. \ref{pb2} is then used, together with the minimum condition  $\partial V/\partial y_{n}= 0$, to determine the	force $F (n_{thresh})$ that is necessary to keep the sequence unzipped up to base pair $n_{thresh}$,
\begin{gather}
F (n_{thresh}) =  2aD e^{-u (n_{thresh} \mid 0)} (1-e^{-u (n_{thresh} \mid 0)}) \nonumber \\
+\frac{K}{a}(u (n_{thresh} \mid 0)-u (n_{thresh} \mid 1)).
\label{pb7}
\end{gather}
When substituting in the above equation the typical values for the parameters of the Peyrard-Bishop model ($D=0.063 eV, K=0.025 eV $\AA$^{-2}$, and $a=4.2$ \AA$^{-1}$ \cite{singh2005} - so that $A=88.9056$) we find that there is a high force barrier for opening the first base pair, see Fig. \ref{Force} (as discussed below, we also observe this barrier in mesoscopic and atomistic simulations). The position and height of the barrier can, in principle, be obtained analytically as a function of $D$, $K$ and by searching for the maximum of $F$ in Eq. \ref{pb7}, but the final expressions are too long and tedious to be reproduced here. Numerically, the force threshold is 218.69 pN. The validity of Eq. \ref{pb7} was checked by integrating numerically Hamilton's equations of motion. This led to a force threshold within 1 pN of the value derived from Eq. \ref{pb7}.
The force-separation curves obtained through the three methods are displayed in Fig.  \ref{Force}. The results obtained using the Peyrard-Bishop model are plotted with a blue dotted line, those obtained from all atom simulations with a red line for fully paired first base-pair, the orange dashed line atomistic simulations for 10$\%$ frayed and 90$\%$ paired first base-pair, the light brown dashed line atomistic simulations for 30$\%$ frayed and 70$\%$ paired first base-pair, and those of the coarse-grained simulations with a magenta dotted line. Moreover, the figure also displays results of coarse-grained simulations for two variations of the DNA sequence: in the first one, the first base pair has been changed from CG to AT (black dotted line), while in the second one the same substitution has been done for the second base pair (magenta dotted line). We note the presence not only of the initial high-force peak to separate the first base pair, but also the presence of an echo, a second peak of a relatively smaller force (appprox. 50 pN, but still larger than the ``bulk" separation force). Finally, we also show, for comparison, the result of Ref. \cite{cocco2} as a green dashed line. In this work the authors develop a semi-microscopic model for the binding of the two nucleic acid strands which also predicts the presence of a high energetic barrier for DNA mechanical unzipping and accounts its origin to the higher rigidity of the double helix as compared to the single DNA strand.

In the coarse-grained and all-atoms simulations, \textit{two} force peaks are also observed at small separations:  the first one is very sharp and occurs at the beginning of unzipping. Its magnitude varies between 132 and 219 pN depending on the model used. This range is in agreement with the values predicted by other theoretical work \cite{cocco2, santosh2009}. It is noteworthy that, in the coarse-grained simulations displayed in Fig.  \ref{Force}, the second peak for for the two dodecamers that have CG base pairing at position 2 the peaks are identical, while for the dodecamer that has AT base pairing at the second position the peak is lower and occurs at slightly smaller interstrand displacement.  Also within reason is the fact that when the \textit{first} unzipped base pair is the same (CG) in two different dodecamers (i.e., in which the first basepair stacks on and AT vs a CG base pair) in the coarse grained simulations, the force profiles are nearly identical for the first peak (see Fig.  \ref{Force}). 

The observation of these two force peaks is explained by the free energy plot in Fig. \ref{FreeEnergy}, which has two local minima (with steeper slopes, yielding the force peaks) at the same separation distances at which the force peaks are observed. 
The inset to Fig. \ref{FreeEnergy} shows that the conformational entropy part of the total entropy contribution to the free energy well around the first base pair for the fully paired first base-pair (black line in Fig. \ref{FreeEnergy})) has a strong contribution, which accounts in part for the high force peak.  These small inter-strand separations (certainly for the first peak and most likely also for the second one) are likely below the minimum resolution one can use to investigate them through typical AFM experiments \cite{albrecht, krautbauer, strunz}. Such peaks are yet to be observed experimentally in singe molecule experiments as increased force-distance resolutions become available. As an encouraging alternative, unzipping experiments  have already recorded pausing events whose magnitude may well be associated with the overcoming of these energy barriers \cite{danilowicz2}.  

An interesting feature is the second force-peak, located at the larger separation associated with the second base-pair rupture. It is weaker ($\sim 50$ pN) than the first peak \textit{and} wider, and reminiscent of the unzipping ``echoes" in Ref. \cite{danilowicz2}.  For even larger separations (echo diminishing, see above)), the force tends towards a constant value (the so called ``critical force," in the large scale unzipping studies), which is the force needed to keep the two DNA strands  separated. It has been shown experimentally \cite{essevazroulet} that this force is constant for homogenous sequences and fluctuates if the sequences are inhomogeneous. Our values for these long-scale separation forces are close to 20 pN, which is within the range of  measured values \cite{danilowicz2, rief}.  Of note is also the fact that, while both the semi-microscopic model of Ref. \cite{cocco2},  as well as the analytical PB treatments both lead to the appearance of a first peak, neither feature a second peak, which is indicative of the fact that they are, in effect, local models. The second peak is observed in both our atomistic and coarse-grained studies because they involve longer range interactions encompassing more degrees of freedom, hinting at a more nuanced picture for the balance of forces at play at the end of DNA duplexes.

The fact that this high barrier for initiation of unzipping is observed in all descriptions gives, on one hand, information about its origin, but it is also a validation of the coarse-grained models. Moreover, by observing the gradual increase in the number of interactions included in these models, we can assess which are the main contributors to the observed force peaks. The Peyrard-Bishop model contains only two terms, namely stacking between consecutive bases on the same strand and pairing between complementary bases on opposite strands. Therefore, the high force needed to initiate DNA unzipping has to stem from to the need to overcome these two types of interactions in a manner that depends on whether the terminal or bulk base pairs are broken. This hypothesis is also confirmed by the fact that, when changing a base pair from CG to AT the magnitude of this force decreases accordingly (a GC base pair contains three hydrogen bonds, while an AT has only two, making it easier to break). Also in common, both coarse grained and atomistic models showed that the base pairing opening pathway was towards the major groove (see Movie in Supplemental Information), which is in accord with previous studies showing this direction as being more favorable \cite{RamsteinL88, BanavaliM02, GiudiceVL03}.  

\begin{figure}[H]
\vspace*{0pt}
\centering{\includegraphics[width= 0.5 \textwidth]{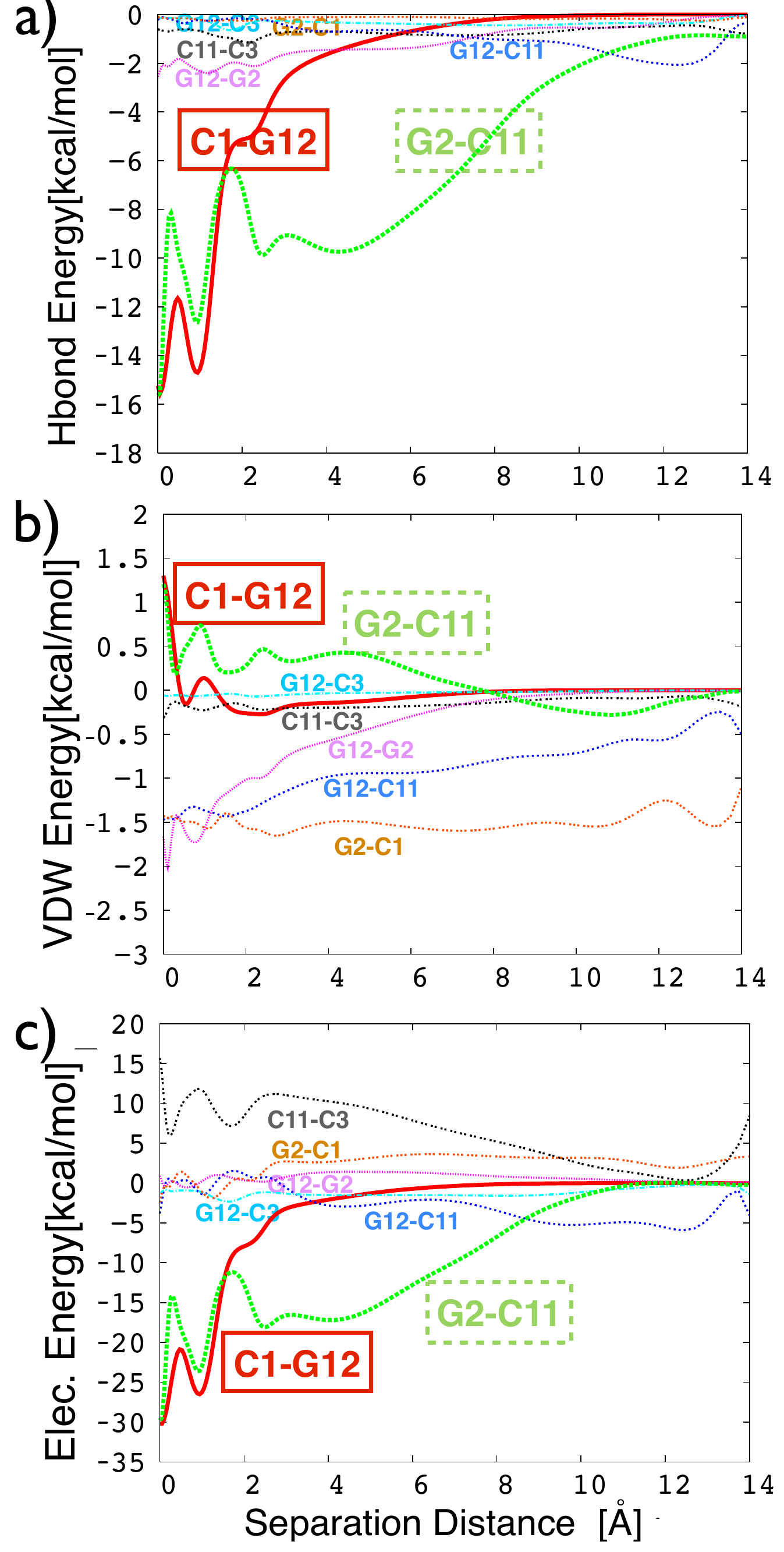}}
\caption{Select terms of the all-atom Hamiltonian as a function of separation distance, averaged over all simulation windows. {\bf a)} Native hydrogen bonds energy between the first and second base pairs (the solid red and dashed green lines, respectively), and some non-native hydrogen bonds energies. {\bf b)} van der Waals term between the first base pair and different bases on the same DNA strand and/or complementary DNA strands.  {\bf c)} Electrostatic interactions between the first base pair and different bases on the same DNA strand and/or complementary DNA strands. }\vspace*{5pt}
\label{allenergy}
\end{figure}

\begin{figure}[H]
\vspace*{0pt}
\centering{\includegraphics[width=0.5 \textwidth]{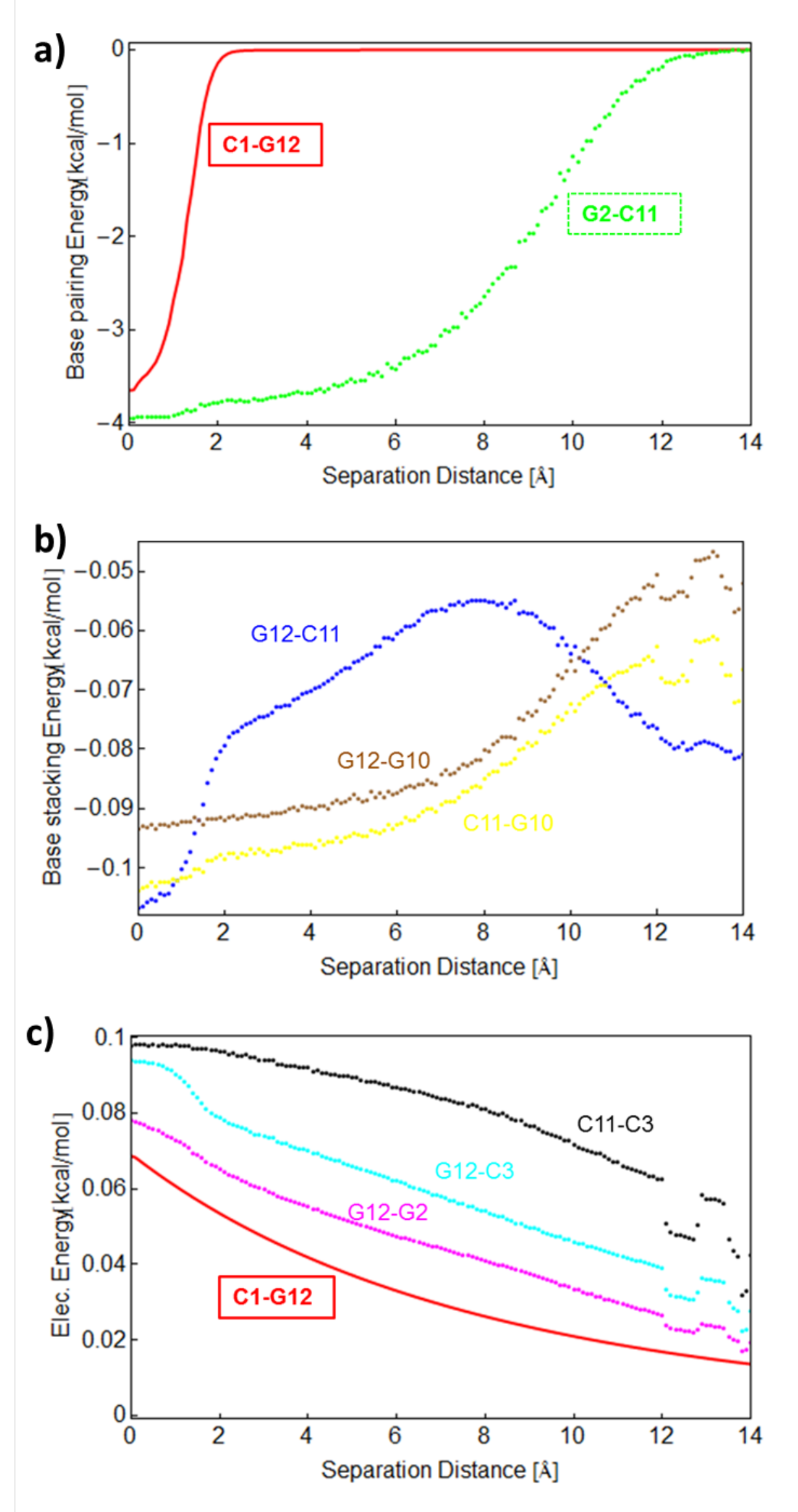}}
\caption{Select terms of the coarse-grained Hamiltonian as a function of separation distance, averaged over all simulation runs. {\bf a)} Pairing energies between the first and second base pairs (the solid red line and the green dots, respectively).{\bf b)} The base stacking energies between first and second and first and third bases on one of the two DNA strand (the curves for the second strand are similar) {\bf c)} Electrostatic interactions between the first base pair and different bases on the complementary DNA strands.}\vspace*{5pt}
\label{allenergy2}
\end{figure}

In order to further check our hypothesis, we plotted separately some of the energy terms for both the all atom and coarse-grained models as a function of the separation distance. They are displayed in Figs. \ref{allenergy} and \ref{allenergy2} for the all atom simulations and the coarse-grained model, respectively.  The first panel depicts the base pairing energy. Our coarse-grained model does not account for non-native pairing so we only plot the energy of the first base pair in red and that of the second one in dotted green. It can be seen clearly that there is a sharp increase in energy at a separation distance corresponding to the first and second force peaks, respectively. Moreover, the slope corresponding to the separation of the second base pair is weaker, thus accounting for the smaller magnitude of the second peak. This is confirmed by the hydrogen bonding energy terms in the all-atom model, which display similar tendencies (see also distance dependence of hydrogen bonding in Supplementary Fig. 1). For the all-atom simulations we also plot some of the non-native pairing terms. As expected, these terms display only small variations, suggesting that their contribution to the force peak is minor. The other panels of Figs. \ref{allenergy} and \ref{allenergy2} display the van der Waals and electrostatic terms, separately (Fig. \ref{allenergy}-b, c) and stacking (Fig. \ref{allenergy2}-b) and electrostatic (Fig. \ref{allenergy2}-c) interactions for the coarse-grained model, for several combinations of bases, either on the same strand or on different strands. 

\begin{figure}[H]
\vspace*{0cm}
\centering{\includegraphics[width=0.8\textwidth]{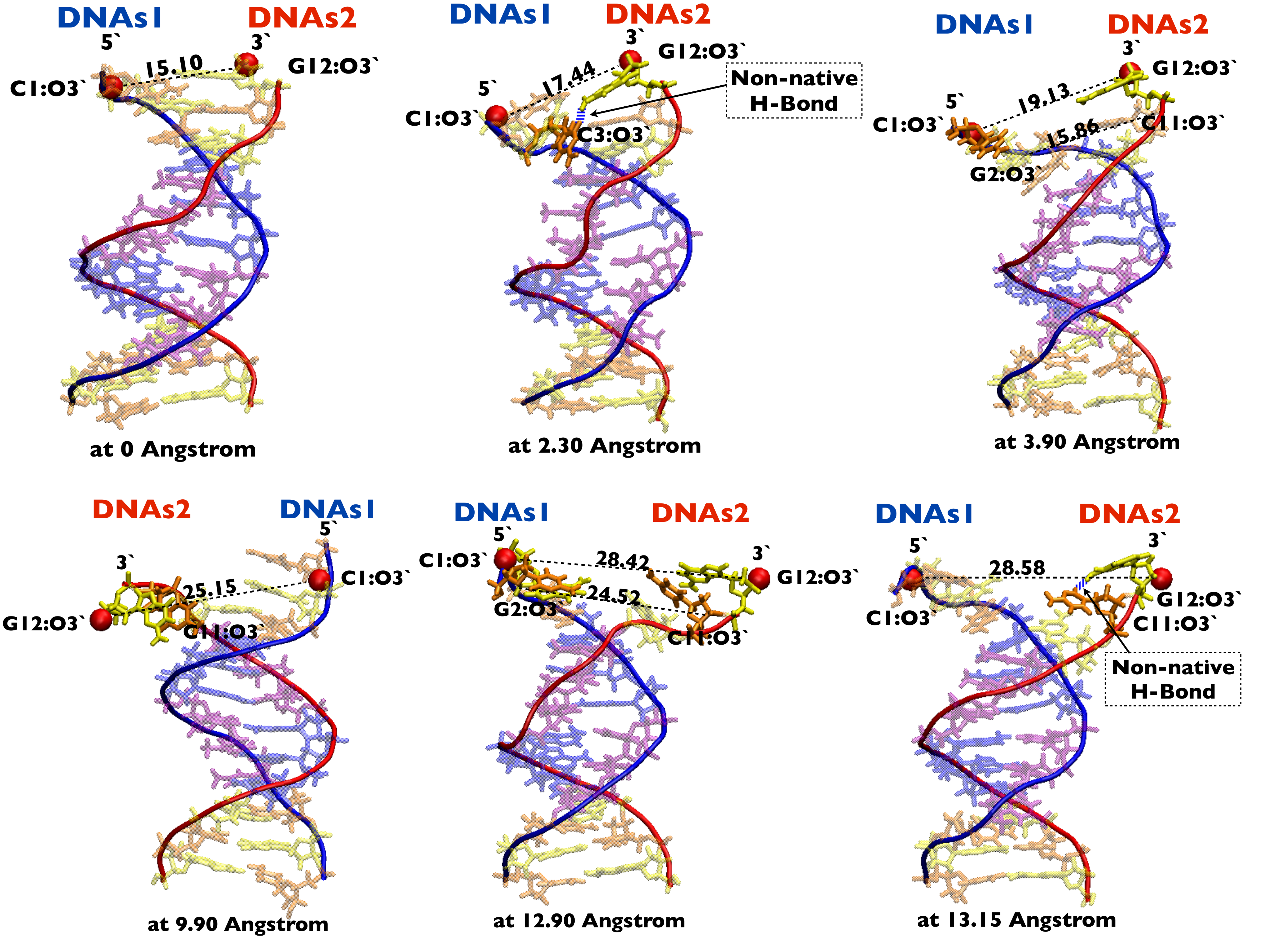}}\vspace*{5pt}\\
\caption{Snapshots of some of the intermediate structures that form while unzipping DNA, extracted from our atomistic simulations using VMD. Transient bonds that form during unzipping are shown with a dashed blue line. Nucleotides G, C, A and T are given in colors yellow, orange, blue and magenta, respectively. The two DNA strand backbones are shown in blue and in red. }
\label{dna_unzipped}
\end{figure}

At this point, comparison between the two types of representations becomes more tedious, because of the simplifications made in the construction of the coarse-grained model.  Firstly, coarse-grained electrostatic interactions only act between phosphates and only have a repulsive part, as can be seen in Fig. \ref{allenergy2}c. They essentially contribute to prevent different DNA segments from overlapping, but they also increase the rigidity and the persistence length of the backbones.  The stacking interactions could be compared to the sum of the van der Waals and electrostatic terms for the CHARMM force field. Both representations predict that the variation of these energy terms is rather small compared to that of the pairing/hydrogen bonding terms. Nevertheless, there is a stacking barrier at the separation distances where the two peaks occur (Fig. \ref{allenergy}b), which shows that they do contribute to the high force needed to initiate unzipping. This is also seen from the intermediate atomic structures displayed in Fig. \ref{dna_unzipped}, which depicts six snapshots of some of the conformations that DNA takes during unzipping, as extracted from the umbrella sampling MD simulations and plotted using VMD. The first panel shows the equilibrated sequence and the red dots show the points where the separating force is applied. The second conformation corresponds to roughly the same separation distance where the force peak occurs. Note an intra-strand ``bond" between the first two bases on the second strand (G12-C11). This is also confirmed by the plot of the energy (van der Waals term) in Fig. \ref{allenergy}-b and the stacking energy in Fig. \ref{allenergy2}-b.  This bond is broken at larger separations (the second configuration), where a transient across-strand bond between G12 and C3 (i.e., a ``non-native" H-bond) is formed, and then again transiently reforms.  Moreover, this also happens for several other non-native H-bonds (as seen in the fourth and sixth panels of Fig. \ref{dna_unzipped}). In the fifth and sixth panels, the second base pair is already opening, but now intra-strands bonds between bases seem to have recovered, as also suggested by the van der Waals G2-C1 and G12-C11 energy terms plotted in Fig. \ref{allenergy}-b.  More generally, we observe that these non-native interactions have either a ``hydrogen bond" or electrostatic character, rather than a van der Waals interaction. It is however difficult to evaluate how much these changes in the interactions along one strand could ultimately contribute to a diminution of its rigidity.  A similar trend is observed for the plots of the stacking energy in the coarse-grained simulations (see Fig. \ref{allenergy2}-b, where the energy terms are only shown for one DNA strand, but are similar for the other one): the energy for the G12-C11 interaction increases at higher separations and then decreases again. It is fair to admit that our coarse-grained model is not detailed enough to capture the change in the nature of the forces determining the interactions.  However, it can be observed from visual inspection of the atomic structures in Fig. \ref{dna_unzipped} (see also Movie in SI), and also from coarse-grained calculations (data not shown) that some of the bases rotate when opening, as suggested by the simulations of reference \cite{santosh2009}. That work also suggested the presence of a torsional barrier to unzipping, which is confirmed in our atomistic simulations (see Supplementary Material).

Taken together, our results suggest that the occurrence of a force barrier when opening the first DNA base pair has two main causes: the breaking of hydrogen bonds between the first bases on each strand, and that of stacking interactions between these bases and their nearest-neighbor along the same strand. When the sequence is completely zipped, there are few fluctuations in its conformation and this is seen in the high forces needed to break the hydrogen bonds. Once the bonds forming the first base pair are broken, the bases have access to more configurations (as evidenced by the higher entropy, see inset to Fig. \ref{FreeEnergy}), fluctuations increase (the chain is also more flexible) and the second base pairs are easier to open. This phenomenology can also be used to describe the microscopic origin of the cooperativity manifested by statistical models. It also agrees with previous simulations of DNA base pair opening which suggest that a strictly local model of the opening of DNA base pairs would not hold \cite{GiudiceVL03,BanavaliM02}. An additional factor to consider is the contribution to the force from the solvent, which was shown to be a major determinant for the dsDNA helical conformation \cite{MaffeoYCWLA14}. Moreover, the interpretation of the initial barrier as having an important contribution from stacking interactions within the same strand (in addition to the breaking of hydrogen bonds) is in line with earlier calculations on the cost of unstacking  \cite{NorbergN95, NorbergN95a}, which showed a 2-4 kcal/mol barrier before the bases become independently solvated at about 2A increased separation

The presence of such a high force barrier involved in DNA mechanical unzipping has already been discussed in the literature, but the presence of a second peak has, to our knowledge, not been previously reported. This second peak is especially well observed in the atomistic simulations, because of their high resolution. We moreover expect that, if these simulations would be run for a higher separation and for a longer sequence, a third peak corresponding to the opening of the third base pair could be observed. Although our coarse-grained simulations were run until the separation distance has reached 50 \AA~we do not observe other force peaks because at higher separations denaturation will proceed in a very fast and irregular manner, with significant noise. This is confirmed by the plot of the number of open base pairs as a function of separation: it shows a large increase at distances higher that 20 \AA, similar to the phase transition that has been observed for larger sequences  \cite{peyrard2004}.  We however ran a set of coarse-grained simulations on a homogeneous CG sequence of similar length, and there we do observe the occurrence of a third peak at a separation distance of $\sim 18 $\AA, which is smaller and wider than the second one (data not shown). Our findings are supported by an earlier atomistic simulation study of  DNA mechanical denaturation using the AMBER force field  \cite{santosh2009}. There, the authors applied an increasing force perpendicular to the helix axis, and plotted the separation and the number of open base pairs as a function of the force, for various temperatures. For 300 K, which is closest to the temperature we use, they observe that the base pairs start to separate at a critical force of 237 pN, after which unzipping progresses through jumps and pauses.  We note, in passing, that this involves and out-of-equilibrium situation, and the use of a different simulation protocol, which does not use the umbrella sampling restraints in our simulations. Similar discrepancies between the experimental force and the rapidly pulling force in simulations was previously discussed in studies of the end to end stretching of DNA \cite{HarrisSL05}.

As discussed in the Introduction, experimental studies have established the fact that DNA ends fray in solution - that is, that the terminal base pair can open spontaneously due to thermal fluctuations. For a sequence with a GC terminal base pair, the fraying probability is around 10\%. We accounted for this in our work by running an additional set of simulations in which the first base pair is initially open (frayed) and then computing a weighted averaged force profile between the two configurations (paired and frayed) with weights that correspond to the experimental probabilities of the open and close state. We also point out that the requirement in statistical mechanical models for a large force to initiate unzipping is not in conflict with the observations of DNA terminal base-pair fraying. For example, using Langer barrier-crosssing theory, Cocco \emph{et al} \cite{cocco2} have shown that there occurs a fluctuation-assisted crossing of the free energy barrier for opening corresponding to this force.

It is instructive at this point to make a comparison between the resolutions and accuracies of the two types of simulations used in this work. The fact that they both manage to capture the position of the two force peaks proves, on one hand, the robustness of the coarse-grained model and, on the other, the entropic-well origin derived from the atomistic representation. The magnitude of the critical force is different in all descriptions used herein, and its value actually decreases when the resolution of the model is increased, suggesting that the coarse-grained models would tend to overestimate this force. One would have expected an opposite trend, since in the simpler model there is a smaller number of configurations available to the system, and thus reduced entropy. On the other hand, in the all atom case one captures more intermediate states with energy very close to that of the fully unzipped base pair, making the transition less abrupt. Although both types of models manage to describe the main phenomena at similar accuracy, all atom simulations bring more information on the details of the intermediate states that occur during unzipping and on their dynamics. Some aspects, like the various types of interactions that form between neighboring bases along the same strand or non-Watson-Crick inter-strand H-bonds can only be captured by atomistic simulations. An ideal method would be a combination of the two: the coarse-grained models would allow the simulations of larger durations, and the more interesting events could then be simulated in more details using atomistic force fields.

\section{Conclusion}

We used molecular simulations at two resolution levels and an analytical (Peyrard-Bishop) model to provide a detailed study of the onset of DNA mechanical denaturation. Our results bring new information on the transient interactions that occur during this process. We observe a large force peak at $\sim2$ \AA~separation and a second, smaller peak at distances ranging between 8 and 12 \AA. We predict that the force peaks on the profile will continue, but with lower values, for the opening of the following base pairs (a third small peak is seen in the coarse-grained simulations, data not shown) but will become indiscernible due to an increase in the ``signal-to-noise" ratio.

To understand the origin of these force peaks, we have computed free energy profiles and we have further analyzed conformational entropy contributions, hydrogen bonding interactions (for both native, i.e., canonical Watson-Crick pairing, and for non-native connections) and stacking interactions of the first few bases at the end of the DNA molecule where the unzipping force is applied. We observe secondary contributions to the force peaks from entropic effects associated with the other types of interactions within the DNA sequence. The essential feature leading to the presence of the initial large force peak(s) comes from analyzing the potential of mean force. The first well is narrower than the second and the subsequent ones because of less entropy (fewer states in the reaction coordinate), hence it is steeper, leading to a larger slope. The force needed for unzipping  is thus higher because of the lower entropy of the chains zipped up for the first base pair as opposed to the second one, and this effect diminishes in ``ripples," or echoes, as the separation between the strands increases. Eventually, when a significant portion of DNA is in single-stranded form, separation forces drop to the 12-20 pN limit observed in the experiments, as base-pair separation becomes akin to melting, which is known to be driven by fluctuations and therefore strongly depends on the conformations available to the now-floppier single-stranded force handles.

The observed higher forces for unzipping initiation relative to the forces needed in single molecule pulling point to a  difference in behavior in boundary vs. bulk pairs. While to date we are not aware of any direct experimental probing to confirm or disapprove the presence of  the large forces needed for initiation, indirect confirmation may exist. For example, proton exchange has been used to probe base-pair opening kinetics in 5'-d(CGCGAATTCGCG)-3' and related dodecamers \cite{Moe1990821}. The enthalpy changes for opening of the central basepairs are correlated to the opening entropy changes. This enthalpy-entropy compensation minimizes the variations in the opening free energies among these central basepairs. Deviations from the enthalpy-entropy compensation pattern are observed for basepairs located close to the ends of the duplex structure, suggesting a different mode of opening for these basepairs. It is possible that the difference in unzipping the first base pairs revealed in our work could be a manifestation of this difference in opening modes observed in the NMR data. 
	
Longer simulations at the actual separation forces with a variety of atomistic DNA force fields or more sophisticated sampling schemes of the actual kinetics of the transition \cite{NummelaA07}, together with new experimental techniques developing increasingly in resolution (such as force clamp spectroscopy \cite{FernandezL04} or nanopore unzipping technologies \cite{McNallyWM08}) may give more insight into the sequence dependent thermodynamics and kinetics of Watson-Crick \cite{nikolova2012} and alternative \cite{NikolovaKWOAA11} base-pairing phenomena.

\section*{Author Contributions} A.M., A.M.F. and I.A. designed research; A.M., A.M.F., E.B. and J.W.  performed research; A.M.F., M.J. and I.A. contributed analytical tools; A.M., A.M.F. and I.A. analyzed data; A.M., A.M.F., E.B., I.A. wrote the paper.

\section*{Acknowledgments}
IA acknowledges support from NIH (grant 5R01GM089846) and the NSF (grant CMMI-0941470). A. M. F. thanks the MPG for support by a postdoctoral grant through the MPG-CNRS GDRE  ``Systems Biology." We also acknowledge NSF Grant CHE-0840513 for the computational resources used for our calculations on the Greenplanet cluster at UC Irvine. 

\section*{Supplementary Information} Several other analyses were performed, gauging the force components and the torque on the terminal base pairs upon unzipping and an additional hydrogen bonding analyses; figures and discussion are included as Supplementary Material. We also uploaded a Supplementary Movie showing the unzipping of the first two base-pairs from the umbrella sampling trajectories. The movie was made taking  20 frames from each umbrella sampling window and used VMD. Nucleotides G, C, A and T are given in colors yellow, orange, blue and magenta, respectively. The two DNA strand backbones are shown in blue and in red. The C1:O$3'$  and G12:O$3'$ atoms that are pulled apart are shown as red balls and their distance is marked with black dotted lines. The first two base-pairs are in opaque and the rest of DNA in transparent representations.

\vspace*{4pt}

\bibliography{unzip19mod_5}
\bibliographystyle{unsrt}

\end{document}